\documentclass[aps,prb,superscriptaddress,twocolumn]{revtex4}
\usepackage[dvips]{graphicx}
\usepackage{amsmath}

\begin{document}

%\preprint{}

\title{Electronic resonance states in metallic nanowires during the
breaking process simulated with the ultimate jellium model}


% \author{E. Ogando,N. Zabala\label{ele}\label{dipc}, T. Torsti\label{fin} and M.J. Puska  }
\author{E. Ogando} \email{eoa@we.lc.ehu.es} \affiliation{Elektrika eta
Elektronika Saila, UPV-EHU 644 P.K., 48080 Bilbo, Spain}
\author{T. Torsti} \affiliation{Laboratory of Physics, Helsinki
University of Technology, P.O.Box 1100, FIN-02015 HUT, Finland}
\affiliation{CSC - Scientific Computing Ltd., P.O. Box 405, FIN-02101 Espoo, Finland}
\author{N. Zabala} \affiliation{Elektrika eta Elektronika Saila,
UPV-EHU 644 P.K., 48080 Bilbo, Spain} \affiliation{Donostia
International Physics Center (DIPC) 1072 P.K., 20080 Donostia, Spain}
\author{M.J. Puska} \affiliation{Laboratory of Physics, Helsinki
University of Technology, P.O.Box 1100, FIN-02015 HUT, Finland}


\date{\today}

\begin{abstract}
We investigate the elongation and breaking process of metallic 
nanowires using the ultimate jellium model in
self-consistent density-functional calculations of the electron
structure. In this model the positive background charge
deforms to follow the electron density and the energy minimization
determines the shape of the system. However, we restrict
the shape of the wires by assuming rotational invariance about the 
wire axis. First we study the stability of infinite wires and show that 
the quantum mechanical shell-structure stabilizes the uniform
cylindrical geometry at given magic radii.  
Next, we focus on finite nanowires supported by leads modeled
by freezing the shape of a uniform wire outside the constriction
volume. We  calculate the conductance during the
elongation process using the adiabatic approximation and the WKB
transmission formula.  We also observe the correlated 
oscillations of the elongation force. In different 
stages of the elongation process two kinds of electronic structures 
appear: one with extended states throughout the
wire and one with an atom-cluster like unit in the constriction and
with well localized states. We discuss the  origin of these structures.
\end{abstract}

%\keywords{}

\maketitle


\section{Introduction}

% What is the meaning of ``topical'' here ?
% Maybe this first paragraph is still a bit wavy ?
The miniaturization of the electronics components is of great %topical
importance in the development and improvement of new devices for
applications  in a wide number of fields. Although the laws of Nature are
the same for macroscopic and mesoscopic systems, the miniaturisation
process is achieving the  limit where the quantum behavior of
matter starts to play an important role.

If the size of the system under consideration is only  a few
nanometers, the atomic character of matter emerges and it can not be
considered as a continuum. The regime of quantum behavior is reached 
also if one of the spatial dimensions of the system is reduced down 
to the Fermi wavelength of the conducting electrons. Then, the confinement
splits the continuous electronic band in this direction into a set of
discrete energy levels. In both cases, the 
behavior of the system changes from what is expected from the
macroscopic case.  In metallic nanowires the Fermi wavelength
is of the same order of magnitude as the atomic distance,
and both atomic and electronic discrete character compete and/or
couple, determining the properties of nanowires.

There are many experimental and theoretical works which have gone
deep into  the understanding of the main features of nanowires.
Experimental studies have focused on the investigation of the
% This was ``electric properties''. Why not ``electronic properties'' ?
mechanical and  electronic properties such as force, atomic structures and
conductance,  pointing out the close relation between them. Among
the  experimental setups we want to emphasize the role of the scanning
tunneling microscope\cite{rubio,rubio-bollinger,brandbyge} (STM) 
and the mechanically controllable break-junction (MCBJ) techniques
\cite{yanson1,yanson2,yanson3}. In  both techniques metallic nanowires are
produced by putting two protrusions in  contact and then pulling 
them away from each other over atomic distances. In this process,
a nanowire is  produced which upon pulling is elongated 
and narrowed until it eventually  breaks. These
methods have allowed the study of transport properties  and
stability of nanowires. 

The MCBJ techniques have demonstrated the existence of
electronic and atomic shell structures \cite{yanson1,yanson2,yanson3},
analogous to those found in atomic clusters \cite{brack,heer}. In  these
experiments the conductance has been studied by building  histograms
of the conductance during the breaking process. The results show that
there are conductance values which are much more probable than others. Due
to the relation between the conductance and the radius at the
narrowest part of the nanowire, that  means that there are magic radii
with enhanced stability while other radii are less stable, and
therefore they appear less frequently in the conductance
histograms. The atomic structures of nanowires in the last steps
before breaking have been studied also with these
techniques \cite{rodrigues1,rubio-bollinger,rodrigues2,ohnishi,yanson4}.

The experiments discussed above have been accompanied by supporting theoretical
investigations, which can be split in two groups. The first group
includes classical and {\it ab initio} molecular dynamics simulations,
in which the atomic structure of nanowires is taken into account.
% The starting point is, for example, molecular dynamics simulations. 
These investigations
have been successful in many aspects, {\em e.g.}, showing the 
atomistic mechanisms of the narrowing process (appearance of dislocations, 
order-disorder stages, {\em etc.}) and their link with other measurable 
quantities such as the elongation force or the conductance
\cite{landman,sorensen}. Moreover, from the viewpoint of the present work
we notice the predictions of special atomic arrangements in STM tips 
and nanowire necks
\cite{landman,hakkinen,silva,sanchez-portal,tosatti,barnett}.
The second group of models is more related to properties due to the confinement
of electrons in reduced dimensions, and ignores the atomistic 
structure of matter. In these  calculations analytic approximations as well 
as  self-consistent electronic-structure models have been used,
mainly within the jellium framework. The results obtained with these methods 
are also enlightening, explaining the cohesive and electronic  transport
properties of nanowires, especially in the case of alkali metals with
strong free-electron character
\cite{edu,martti,nerea,yannouleas,yannouleas2}.

The aim of this paper is to simulate the breaking of
nanowires. For this purpose we choose the jellium model and 
the self-consistent electronic-structure calculations within the
density-functional theory. In spite of their simplicity jellium models 
have provided a simple and transparent way to  understand the physics of 
metallic nanowires. More specifically, we use the ultimate jellium  (UJ)
model. This model was first proposed by Manninen \cite{manninen} 
to investigate the structures of alkali metal clusters. It
has been used for the same  purpose also in later
studies \cite{koskinen,koskinen1}. To our knowledge the present work is 
the first time the UJ-model is used to simulate the nanowire breaking. 
In practice, we solve the ensuing Kohn-Sham equations in a real-space 
point grid using the powerful Rayleigh Quotient 
Multigrid\cite{mandel,heiskanen}  (RQMG) method implemented in the program 
package MIKA\cite{tuomas}  (Multigrid Instead of K-spAce). 

Within the UJ approach, the background positive charge density
is fully relaxed in shape and density so that it equals at every point with
the electron density. One can think that this freedom of the positive
background charge mimics the efficient rearrangement and diffusion of ions
at temperatures close to the melting point at which the shell- and
supershell-structure studies by the MCBJ techniques have been performed 
for alkali metals \cite{yanson2,yanson1}. In principle, there is no 
restriction for the geometry of the constriction. This is in contrast
with the previous jellium calculations, which introduced  
{\it ad hoc} shapes for the nanowire. In our model the electrons
themselves acquire self-consistently the shape, which minimizes the
Kohn-Sham energy functional, and carry along the positive background
charge. However, 
in order to reduce computational demands and to highlight the
important phenomena from the complexity of possible solutions,
we restrict the shapes of nanowires to the axial symmetry, 
{\it i.e.} rotational invariance with respect to an axis.

%Transport and cohesive properties are studied during the elongation
%process. 
One of our main results is that in the narrowest part of the
nanowire, electronic cluster derived  structures\cite{barnett,landman}  
(CDS's) appear.  This tendency of electrons to form embedded 
clusters in  the jellium constrictions is analogous to 
the preferred cluster-like
arrangements of atoms in contacts described by first-principles atomistic
calculations by Barnett and Landman \cite{barnett,landman}.
CDS's have later been reported also by other
authors \cite{hakkinen}. The main difference is that in our jellium
model the atomistic character of the previous works is lost and the
electrons alone are responsible for the phenomenon. 
%The jellium model 
%provides electron eigenfunctions by which the localization effects of the
%CDS's occuring in the nanoconstrictions can be easily studied and the 
%conductance of the constriction can be determined.
The single-electron states provided by the jellium model can be studied
in order to gain insight into the localization effects associated with
the CDS. The conductance of the constriction can be estimated either
by counting the bands crossing the Fermi level, or by using the WKB 
formula.

The rest of the  paper is organized as
follows: in Section \ref{Theory} we describe the practical features  of the
UJ-model and the RQMG-method to calculate the electronic structure
during  the elongation process. In Section \ref{Results} we discuss the results
for the electronic  properties. As a starting point, we consider the results 
for infinite wires. Then we focus our attention on the breaking process
of a finite cylindrical UJ nanowire supported by leads. Section 
\ref{Conclusions} contains the conclusions.

\section{\label{Theory}Theory}

\subsection{Jellium models}

The jellium model has been widely used in self-consistent electronic-structure
calculations of nanostructures. It simplifies the problem by replacing the 
ions by a uniform rigid positive charge density background, which globally
neutralizes the electron negative charge. The effective potential 
of the Kohn-Sham \cite{jones} equations is written as (Atomic Hartree units
are used throughout this paper to write the equations)
\begin {equation}
V_{\rm eff}=\int\frac{n_-(r')-n_+(r')}{|r-r'|}dr'+ v_{\rm xc}[n_-(r)],
\end {equation}
where the first term on the right-hand side includes the 
electron-electron and electron-positive background Coulomb
interactions and the
second term gives the exchange-correlation potential within the local
density approximation\cite{perdew-zunger,ceperley-Alder}  (LDA).

Different types of jellium approaches have been introduced.
The simple jellium (SJ) model has the problem that there is only one 
equilibrium charge density, at $r_s\approx 4.18\ a_0$
($n_-=3/(4\pi r_s^3)$) corresponding approximately to the average 
conduction electron density in Na or K metals. This means that for $r_s$
values lower (higher) than $\sim$4.18 $a_0$ the jellium system tends to 
expand (compress). In the SJ-model the electron density has 
the same mean value as the positive background due to the electrostatic forces. 
The SJ-model gives incorrect values for
properties such as the cohesive energy, surface energy and bulk modulus,
due to the trend of the system to compress or expand. To improve
the results corrections can be added to the SJ-model \cite{shore}, 
{\em e.g.} using the so-called stabilized jellium model
introduced by Perdew \cite{perdew} and Shore \cite{shore2}.

In this work we use the UJ-model, the philosophy of which differs
from the stabilized jellium model in that it does not
try to correct the above-mentioned deficiencies of the
SJ-model. The peculiarity of the UJ-model is that the positive charge 
background is allowed to relax. The UJ-model represents the ultimate limit 
in which the positive background is completely deformed to have the same 
density as the electrons locally at every point. In this way, the Coulomb 
term in the potential always vanishes, and in Eq. (1) only
the exchange-correlation term survives. The total energy is then
minimized in the interplay between the exchange-correlation and the
kinetic energies.

One limitation of the UJ-model is that, as in the SJ-model, there is
only  one equilibrium charge density, at $r_s\approx 4.18\ a_0$.  But,
the absence of electrostatic potential disables the  mechanism to keep
the electrons at a given density, and inside the UJ the mean electron density
becomes equal to the equilibrium density.
Another property of the UJ-model, derived also from the absence of
electrostatic potential, is that the shape of the electron density is
to a large extent uncontrollable, and it evolves until the 
ground state is achieved.  This property has been used to study  the most
favorable shapes of simple-metal atom
clusters\cite{manninen,koskinen,koskinen1,reimann}.
In the present work, however, we have to deal with open systems and
we have to impose certain controlling restrictions in order to
model the pulling of the nanowires. The description of the
solutions to these requirements is postponed to Section \ref{UJ}.

\subsection{Numerical methods}

% multigrid (Brandt)
% RQMG (McCormick)
% SRQMG (Heiskanen)
% MIKA (Torsti)
% equations in cylindrical coords
% mixing (note: UJ is easy !)
% Finite temperature. 
% (Kohn-Sham scheme ?)

In Section \ref{uniform} infinite uniform cylindrical wires are studied.
Since these systems are translationally invariant along the wire axis,
the relaxation of the positive background charge and electron density
is limited in the radial direction. Consequently, it is necessary to 
solve numerically only the radial part of the Schr\"odinger equation
(see Zabala {\it et al.}\cite{nerea} for technical details).

For the systems studied in  Sections \ref{stability} and \ref{UJ}, 
however, the translational invariance is not required. But, in
addition to the rotational invariance, periodicity in the axial direction
is assumed with unit cell length $L_{\rm cell}$. 
Thus, the wave functions $\psi$
are indexed by the quantum numbers $m$, $n$, and $k_z$. Here,
$m$ is the angular momentum quantum number 
% and $n$ is related with the 
% number of nodes of the wave function in the radial direction. 
and $k_z$ is the Bloch wave vector along the wire axis. 
With $m$ and $k_z$ given, $n$ enumerates the orthogonal states
in the order of increasing energy eigenvalue. The UJ system is
solved by finding the self-consistent solution to the  
%and in a given effective potential $V_{\rm eff}(r,z)$ they fulfill the 
set of equations
\begin {equation}
\psi_{mk_zn}(r,z,\phi)=e^{im\phi}U_{mk_zn}(r,z),
\end {equation}
\begin{equation}
U_{mk_zn}(r,z+L_{\rm cell}) = e^{ik_zL_{\rm cell}}U_{mk_zn}(r,z).
\label{bloch}
\end{equation}
%\begin{equation}
%\label{kohnshameq}
%-\frac{1}{2}  \left(\frac{1}{r}\frac{\partial}{\partial r} + 
%\frac{\partial^2}{\partial r^2}  - \frac{m^2}{r^2}  \frac{\partial^2}{\partial z^2} + 2V_{\rm eff}(r,z) \right) &  \\
% \times U_{mk_zn}(r,z)  =
%  \varepsilon_{mk_zn}  U_{mk_zn} & (r,z). \\
%\end{split}
%
\begin{widetext}
\begin{equation}
\label{kohnshameq}
-\frac{1}{2}  \left(\frac{1}{r}\frac{\partial}{\partial r} + 
\frac{\partial^2}{\partial r^2}  - \frac{m^2}{r^2}  \frac{\partial^2}{\partial z^2} + 2V_{\rm eff}(r,z) \right) 
  U_{mk_zn}(r,z)  =
  \varepsilon_{mk_zn}  U_{mk_zn} (r,z). 
\end{equation}
\end{widetext}
\begin{equation}
\label{densitydef}
%  n({\bf r}) = \sum\limits_i^N f_{m{\bf i}}|U_{m{\bf i}}({\bf r})|^2,
  n({\bf r}) = 2\sum_{mk_zn} (2-\delta_{0m}) f_{mk_zn}|U_{mk_zn}({\bf r})|^2,
\end{equation}
\begin{equation}
\label{veffdef}
V_{\rm eff}(r,z) =  V_{\rm xc}(r,z) = \frac{\delta E_{\rm xc}[n(r,z)]}{\delta n(r,z)}.
\end{equation}
%\begin{equation}
%\label{vxcdef}
%  V_{\rm xc}({\bf r}) = \frac{\delta E_{\rm xc}[n({\bf r})]}
%      {\delta n({\bf r})}.
%\end{equation}
The effective potential $V_{\rm eff}(r,z)$ equals the exchange-correlation 
potential $V_{\rm xc}(r,z)$. The electron density $n(r,z)$ is obtained by
summing single-electron densities with the occupation numbers $f_{mk_zn}$. 
The degeneracies of the states are taken into account by the factor 
$2(2-\delta _{0m})$ and the occupation
numbers $f_{mk_zn}$  obey the Fermi-Dirac statistics with the Fermi level 
($E_F$) so that the system is neutral. 
A finite  temperature of 1200 K is used to stabilize the solution of the
set of equations.

The Schr\"odinger equation (\ref{kohnshameq}) is discretized on 
a regular two-dimensional $(r,z)$ point mesh. We use standard fourth-order 
central-difference discretizations for the  first and the second derivatives.
The grid is surrounded by a {\it frame} with the thickness of two 
grid points. These {\it ghost points} are necessary for the evaluation
of the derivatives near the edges of the computation volume. 
The wave functions are required to vanish at the ghost points
corresponding to the radial surface of the cylindrical computation volume, 
whereas at the axis the values at ghost points can be evaluated by noting that
$U(-r,z)=(-1)^mU(r,z)$. The Bloch-condition (Eq. (\ref{bloch})) gives the
recipe for obtaining the values at the ghost points of the periodic boundary.

The problem with standard real-space relaxation methods for Eq.  (\ref{kohnshameq})
is the so called critical slowing-down  phenomenon resulting from 
the fact  that at a time they use information from a rather localized region 
of space. As a result of the locality, the high-frequency error, corresponding
to the length scale of the grid spacing, is reduced very rapidly
in the relaxation. However, once the high frequency error has been
effectively removed, the very slow convergence of the low-frequency
components dominates the overall error reduction rate, {\em i.e.} critical slowing-down occurs.
Multigrid methods avoid this problem by treating the low frequency components
of the error on coarser grids, where their wavelength is  comparable
to the grid spacing. 

Applying the multigrid methods to the Schr\"odinger equation is a fairly
complicated task because one has to solve both the eigenvalue and 
the wave function simultaneously -- this makes the problem nonlinear. 
Also, one has to solve several wave functions simultaneously, avoiding
the bottleneck of orthogonalizations as well as possible.
Standard methods based on the full-approximation-storage\cite{brandt2}
method require, that the wave functions are well representable on the
coarsest grid used, implying severe limitations on the acceleration
obtained by the multigrid idea. 
We use the recently developed generalization of the Rayleigh-quotient multigrid (RQMG)
method\cite{mandel,heiskanen} as implemented in the MIKA-package\cite{tuomas}, that
avoids the problems described above. In short, one applies the Gauss-Seidel method 
on the finest grid. On the coarser grids one applies coordinate relaxations
on the functional
%, the sum of the Rayleigh quotient and a penalty functional to insure 
%orthogonality, which is defined on the fine grid only:
\begin{equation}
\label{rqmgneq}
\frac{\langle \psi_{n}\arrowvert H\arrowvert \psi_{n}\rangle}
      {\langle \psi_{n}\arrowvert \psi_{n}\rangle}
 + \sum\limits_{i=1}^{n-1}
    q_i \frac{\left|\langle \psi_i | \psi_{n}\rangle\right|^2}
             {\langle \psi_i | \psi_i\rangle \cdot
              \langle \psi_{n} | \psi_{n}\rangle}.
\end{equation}
This functional, which is actually  defined on the finest grid, is 
the sum of the Rayleigh quotient and a penalty functional, which is introduced 
to insure the orthogonality.
Moreover, the relaxations are performed simultaneously for all wave functions.
See  Heiskanen {\em et al.}\cite{heiskanen} for a more thorough discussion of 
technical details. 

The Kohn-Sham equations have to be solved self-consistently. In other 
words one has to iterate until the  output potential $V_{\rm eff}$
obtained from Eq. (\ref{veffdef}) equals the input potential $V_{\rm eff}$ that is used in 
Eq. (\ref{kohnshameq}). 
% In contrast to typical cases of
% electronic structure calculations, where 
In typical cases of electronic structure calculations, to avoid divergence 
due to charge sloshing,  one uses sophisticated
strategies to construct the input potential for the next iteration as an optimized
mixture of input and output potentials of previous iterations\cite{raczkowski, annett}.
In the UJ-iterations, however, the output potential can be taken directly as the input
potential of the next iteration resulting in a rapid convergence.
This is because of the absense of the
long-range Coulomb interaction, which is the cause of the 
charge sloshing phenomenon.

\section{\label{Results}Results and discussion}

\subsection {\label{uniform}Infinite uniform cylindrical wires}

The main results of this paper concerning the nanowire breaking process are 
discussed in Sections \ref{stability} and \ref{UJ}. As a preliminary work, 
and in order to gain insight into the UJ-model in comparison with the 
stabilized jellium model, we study the stability of infinite 
uniform cylindrical UJ-wires. 
% ************ I moved this to the previous section (Tuomas) *************
%This means that the system is translationally
%invariant along the wire axis and the relaxation of the background charge 
%is limited in the radial direction. Consequently, it is necessary to solve 
%numerically only the radial part of the Schr\"odinger equation 
%(see Zabala {\it et al.}\cite{nerea} for technical details).

\begin{figure}
\includegraphics[bb= 65 235 540 595 ,width=\columnwidth]{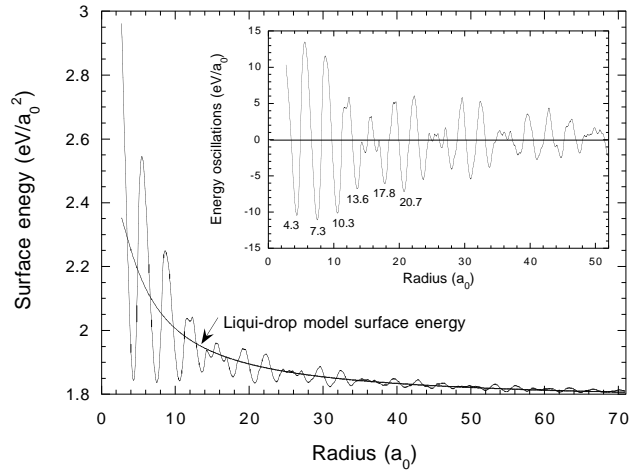}
\caption{\label{ener} Surface energy of infinite uniform cylindrical UJ
nanowires as a function of the nominal wire radius (See text). In the inset the
energy oscillations are shown and the first magic radii are marked.}
\end{figure}

We calculate the surface energy of the nanowires and the
oscillations in the energy per unit length as it was made in 
our previous work describing Al, Na, and Cs nanowires within the stabilized 
jellium model \cite{edu,martti}. The results are shown in Fig. \ref{ener} 
as a function of the wire radius $R$. 
Here the radius is defined as the radius of the positive background charge
in the SJ system with $r_s$ = 4.18 $a_0$ and the same amount of charge
per unit length. In order to separate the energy oscillation from the average 
behavior, the so-called
liquid-drop model \cite{ziesche} is used. In this model the energy of the 
jellium system can be written as the sum of two terms -- one proportional
to the volume and the other proportional to the surface area. 
For the first term, the energy/volume ratio corresponds naturally to 
the homogeneous electron gas\cite{perdew91}  with $r_s$ = 4.18 $a_0$.
%jellium system can be written as the sum of the homogeneous electron gas 
%and the surface energy \cite{perdew91,ziesche}. 
This view has been tested in clusters and
nanowires \cite{nerea,martti,edu} and it describes correctly the mean
energy, {\em i.e.} without the characteristic oscillations due to the
quantum confinement. We fit the self-consistently calculated total
energy per unit length to a liquid-drop model type
function. Then, subtracting this smooth energy function from the total
energy we get the pure energy oscillations, which are shown in
the inset of Fig. \ref{ener}. Note that there are radii for which the energy
is at minimum. They correspond to wires which are more stable than 
wires with slightly different radii and higher
energies. The first magic radii are at $R=4.3$, 7.3, 10.3,
13.6, 17.8, and 20.7 $a_0$. 
We use these radii for the initial uniform wires in the nanowire breaking
simulation in Section \ref{UJ}.
The shell and super-shell structures
studied in previous calculations \cite{martti,edu} are also quite
clear. In comparison with the energy oscillations of Na we
observe that the beat positions are shifted to higher radii. The reason
is that the UJ potential is softer at the surface
than the stabilized jellium potential for Na \cite{edu}.

\subsection {\label{stability} Periodic systems}

Now we change the scheme and allow  the wire to deform also in the
axial direction. However, we impose periodic boundary conditions
with the unit cell length $L_{\rm cell}$ along the wire axis. $L_{\rm cell}$
is thus the maximum perturbation wavelength in our calculation. 
From the liquid-drop model point of view, neglecting the small contribution 
of the curvature energy, the liquid wire attempts to achieve the shape which 
minimizes the surface, and thereby the total energy. Under this assumption 
an infinite periodic liquid wire is a uniform cylinder 
for lengths $L_{\rm cell} < 4.5 R$. For $L_{\rm cell} > 4.5 R$ it deforms trying to
achieve the energetically most favorable state, an infinite chain of spheres. 
However, Kassubek {\it et al.} \cite {kassubek} showed using a semi-classical 
model and perturbation theory that due to the discreteness of
electronic structure the wires with magic radii remain uniform
also at large $L_{\rm cell} / R$ ratios. With
this result they argued that in the narrowing process of an infinite
wire, when the radius is crossing an unstable zone before the next
stable radius is achieved, the wire would spontaneously deform acquiring a
wavy or deformed shape. We corroborate these results
non-perturbatively using the UJ model as follows.

%In this work, we always restrict the nanowire geometry to axial symmetry,
%{\it i.e.} rotational invariance with respect to an axis. Moreover our 
%model systems are either periodic (Sections \ref
%and solve numerically for the wavefunctions of the form $e^{im\phi}U(r,z)$  
%with the MIKA package. Along the wire Bloch boundary conditions are
%imposed. 

\begin{figure}
\includegraphics[bb= 30 168 544 586,width=\columnwidth]{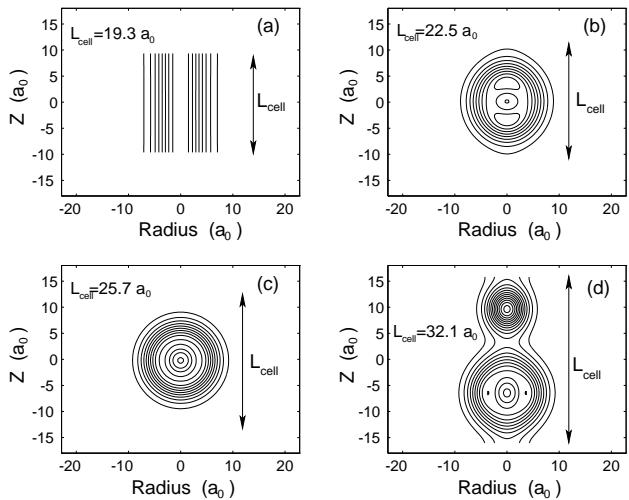}
\caption{\label{periodic} Periodic infinite UJ wire with the nominal radius of 
$R=5.5$ $a_0$ and 6 (a), 7 (b), 8 (c), and 10 (d) electrons
in the periodic unit cell. The figures show the electron density of one unit cell.
The contour spacing is 0.15 times the UJ bulk density value.}
\end{figure}

We choose a certain radius $R$ and solve for the UJ electronic 
(and positive charge density) structures imposing increasingly
longer supercell lengths $L_{\rm cell}$ by increasing the number of
electrons in the cell. Thereby we determine the critical supercell 
length (the wavelength of a perturbation) at which the wire
starts deforming. For magic wires we find no wavy solutions, the wires 
remain uniform. For example, for $R=7.3\ a_0$ the wire is still uniform 
at $L_{\rm cell}/R \approx 36$. The wires corresponding to the radii at the 
maxima of the energy oscillations in Fig. \ref{ener}  are the most unstable 
ones. These wires are uniform up to a critical value of $L_{\rm cell}$ but above 
it they spontaneously deform to a wavy or non-uniform density profile along 
the wire axis. As an example, Fig. \ref{periodic} shows the behaviour of 
a wire with radius $R=5.5$ $a_0$  when the number of electrons in the 
unit cell is 6, 7, 8, and 10 and the unit cell lenght $L_{\rm cell}$ 
increases as 19.3, 22.5, 25.7 and 32.1 $a_0$, respectively. The unit cell with eight
electrons correspond to a magic spherical cluster and that of ten
electrons the pair of magic clusters of eight and two electrons. 
The critical values for the unstable radii of 
$R=5.5$, 8.6, 11.6 and 19 $a_0$ are $L_{\rm cell}/R=$4.1, 3.2, 4.2, and 4.8,
respectively. {\em I.e.}, we obtain values near the classical
value of 4.5. At the unstable radius of $R=15.5\ a_0$ the wire is not deformed
at least up to $L_{\rm cell}/R=10$ (the largest length we have calculated),
probably due to the fact that this radius lies in a beat of the
super-shell structure and it is actually relatively stable. We
start all the calculations with a converged  uniform potential profile 
along the wire axis (See Section \ref{uniform}). In this way we do not 
``add any energy'' to the system when initiating the calculation.
% with more degrees of freedom.
Therefore, if the wire starts to deform in the iteration process the
reason is the disappearance of the local energy minimum.

In addition, we narrow a stable uniform wire by increasing the
length $L_{\rm cell}$ of the periodic cell and maintain the number of electrons
constant. Each elongation step is solved self-consistently until
convergence is reached. We observe that during the first steps the
wire remains uniform but at some point, before breaking into
isolated clusters,  the wire spontaneously
deforms. Thus, we confirm self-consistently and dynamically the
hypothesis by Kassubek {\it et al.}  \cite {kassubek}.

\begin{figure}
\includegraphics[width=\columnwidth]{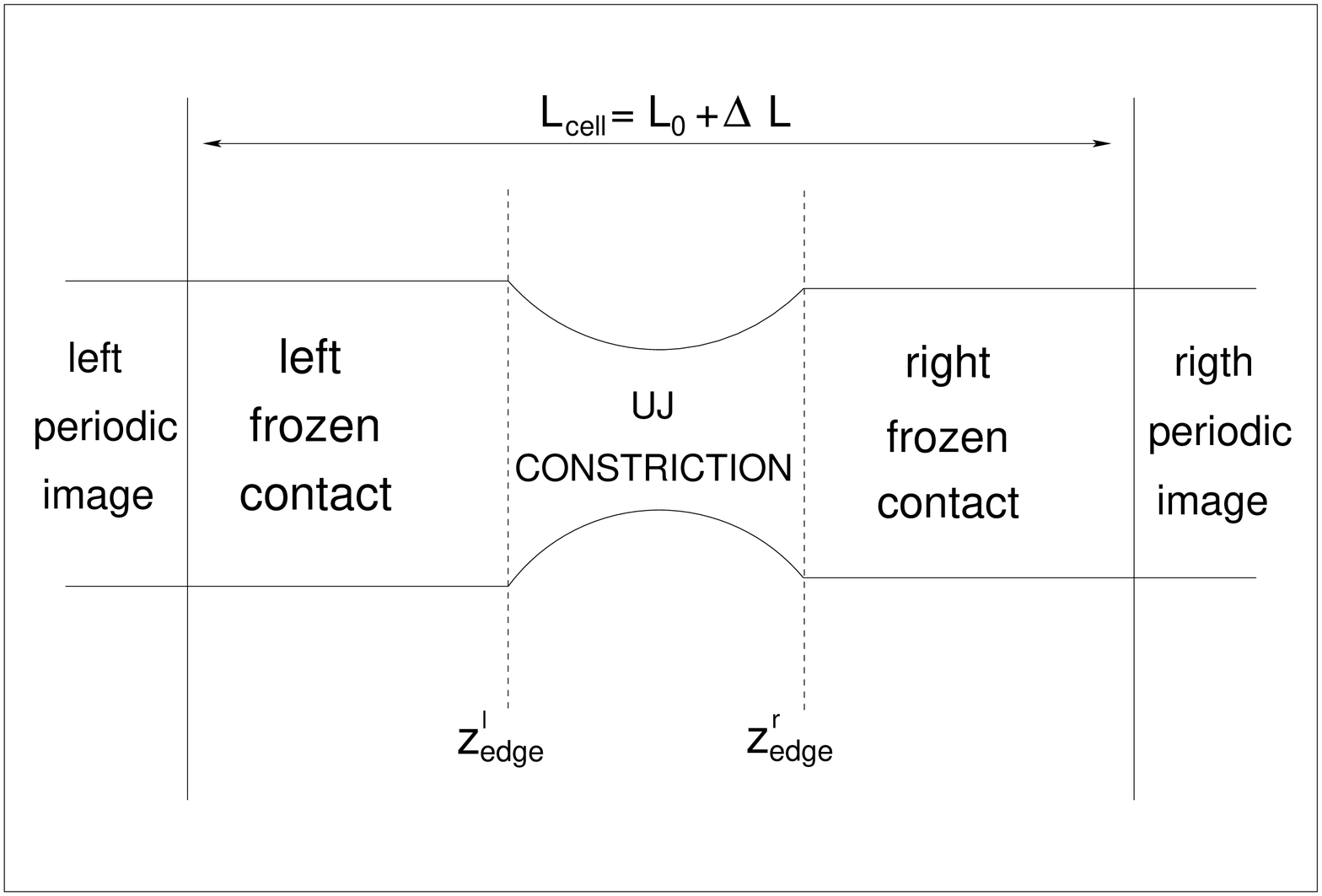}
\caption{\label{sketch} Schematic view of the model system
for simulations of breaking of finite nanowires supported by two leads. }
\end{figure}

\subsection {\label{UJ}Breaking of supported finite nanowires}

%The method described above is not reliable for our purpouses being our
%main aim is to study the formation and evolution of the
%nanoconstriction, 

In order to study the formation and evolution of nanoconstrictions
between two supporting leads we follow the next procedure.
%to build up a nanocontact. 
First, we fix the number of electrons in the periodic supercell and solve 
self-consistently for the electronic structure of a uniform UJ wire
having a stable magic radius. Then, the
potential at both ends of the periodic cell is "frozen". This means
that, although the Kohn-Sham equations are solved in the whole wire,
in these regions the potential is not updated in the
self-consistency process. The
function of this "frozen" part is to  emulate the lead parts where
ion rearrangement does not occur as efficiently as at the constriction. 
In our calculation, these leads serve as handles to grab the 
UJ and pull it. The rest of the wire, the UJ at the middle part of the
supercell, is the place where the wire will stretch. A sketch of the 
configuration is shown in Fig. \ref{sketch}. A sharp change in the 
potential between the constriction and the leads turned out to 
cause difficulties in numerical calculations. Therefore, we smooth out 
the potential at the left edge using the form 
\begin{equation}
F(z-z_{\rm edge}^{\rm l})V_{\rm frozen} + F(z^{\rm l}_{\rm edge}-z)V_{\rm UJ}, 
\end{equation}
where F is a Fermi function with half-with of $0.5\ a_0$, $V_{\rm frozen}$ 
is the "frozen" potential and $V_{\rm UJ}$ is the self-consistent UJ 
potential. For the right edge an analogous mixing is used. The main 
properties of the nanowire will not depend on the particular choice 
of this matching because the physical features are determined by 
the narrowest part of the constriction.

%With this procedure we can say that the wire has been
%separated into  the leads and the constriction.  When the periodic
%cell is elongated the electron density in the leads is attracted 
%by the "frozen" potential and it remains more or
%less unchanged. The leads hold the electron density in the deforming 
%constriction.  Far from the constriction, at the cell boundaries, the 
%structure of the infinite wire is recovered.

We perform simulations starting with radii between $7.3\
a_0$ and $20.7\ a_0$ and changing the number of electrons initially
in the constriction. The elongation of the wire is
made in steps of about one atomic unit, and always starting from
the previous converged density, so that the grid spacing of the
point mesh is increased to enlarge the cell. 
%The interaction between the periodic
%neighbor images of the constrictions could add interferences in the
%calculated properties \cite{stafford-comment}. 
In order to overcome the interactions between the constriction and
its periodic replica  \cite{stafford-comment}, we choose the 
length of the lead part to be 6 Fermi wavelengths 
%between the periodic images of the constrictions 
($\lambda_F=13.7\ a_0$). Throughout the rest of
the paper we will use $\Delta L$ for the elongation; $\Delta
L=0$ for the first step.

\begin{figure}
\includegraphics[bb= 42 80 616 536,width=\columnwidth]{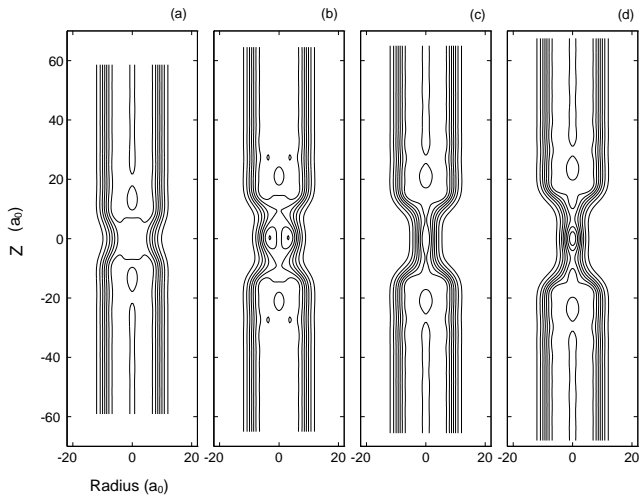}
\caption{\label{rho} Supported UJ wire. The UJ constriction contains
eight electrons. Density contour plots for four different
elongation lengths: $\Delta L=7.9$ $a_0$ (a), 19.8 $a_0$ (b), 
20.8 $a_0$ (c), and 25.8 $a_0$ (d) are shown. The snapshots in (b) and (c)
are from consecutive self-consistent calculations and the snapshot (d) 
is the last step before the nanowire breaking. The contour spacing
is 0.15 times the mean UJ bulk density value.}
\end{figure}

In Fig. \ref{rho} we show snapshots of the electronic
density for a wire with the starting radius of 10.7 $a_0$. 
The UJ-part corresponds to eight UJ-electrons placed initially in the 
neck region. Electrons are
free to move inside or outside the leads, depending on the
requirements of the self-consistent solution. However, there are always
about eight electrons in the constriction. Although this is one of the
smallest wires we have calculated, it shows all the main features observed 
when simulating also larger wires.

If the breaking of an UJ nanowire would happen as for fluid 
between the leads, the electron density should evolve forming a catenoid-shaped
surface. Similar shapes (like hyperbolic \cite{torres}, parabolic
\cite{yannouleas}, cosine \cite{stafford}, {\it etc.}) have been used before
to model the nanoconstriction in simple free-electron or jellium 
simulations. The main results, when the comparison is possible, have been 
essentially the same irrespective of the actual shape. In Fig. \ref{rho}(a) the
electron density is shown after the elongation of $\Delta L= 7.9\ a_0$.
The catenoid-like density profile appears as expected for a
classical fluid. When we continue elongating the nanowire the shape of
the electron density changes dramatically from the classical
one. If the distance between the leads is short the electrons 
are strongly trapped at the narrowest part 
and they do not have much freedom in the rearrangement process.
When the length of the constriction is large enough, the
electrons have more space and freedom to achieve different types of
energetically preferred shapes.

\begin{figure}
\includegraphics[clip,bb= 75 55 490 790 ,width=\columnwidth]{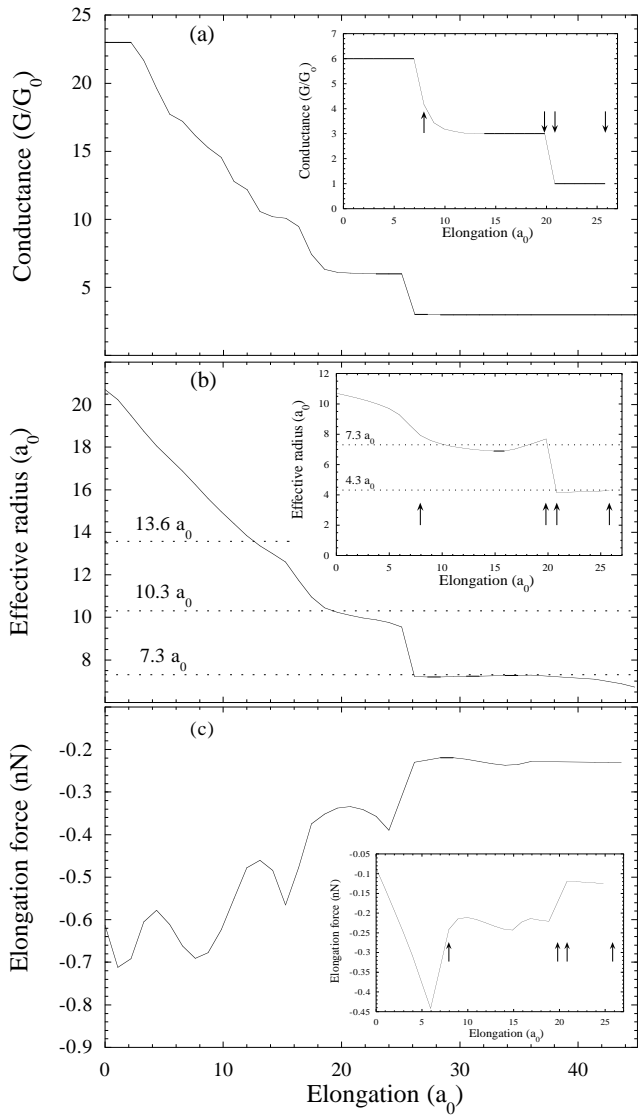}
\caption{\label{force} Main figures: conductance, effective
radius at the constriction, and elongation force for a wire with
initial radius $R= 20.7\ a_0$ and about 60 UJ-electrons in the
constriction. Insets: the same quantities for the wire
in Fig. \ref{rho} with initial radius $R=10\ a_0$ and 8
UJ-electrons in the constriction. The arrows mark the points were the
density has been plotted in Fig. \ref{rho}}. 
%The arrowheads denote
%effective radii calculated using the electron density at the middle
%of the constriction.}
\end{figure}

In Fig. \ref{rho}(b) $\Delta L=19.8\ a_0$ and the electrons in the
constriction form a CDS. The electron density per unit length has two
minima at both sides of the CDS and there are $7.1$ electrons between
these narrowest cross sections. The embedded cluster reminds  the
closed-shell cluster of eight electrons, but there are some differences.
There are not enough electrons and the symmetry is not exactly
spherical.  It seems that the $p_z$ orbital ($z$ along the cylinder
axis) of the  cluster has disappeared. We will analyse the structure
in more detail below. Figure \ref{rho}(c) shows the next consecutive
elongation step  with $\Delta L=20.8\ a_0$. Note that the CDS
disappears and a sudden change in the mean radius happens. In
fact, the conductance changes simultaneously abruptly from $3\ G_0$ to
$1\ G_0$ (See the inset in Fig. \ref{force}(a)). At this point it is 
also remarkable that the shape of the
constriction  is again far from the catenoid having a constant magic
radius.  Figure \ref{rho}(d) is for $\Delta L=25.8\ a_0$, the last
step before the nanowire breaks. Again a CDS appears during the
elongation from the third to the fourth snapshot. There are  $1.8$
electrons between the two minimum cross sections at both
sides of the CDS. This CDS can be
interpreted as an embedded two-electron cluster. We observe that
the radius of the constrictions is more or less constant with the same
value as in the previous snapshot in Fig. \ref{rho}(c).

At this point we want to focus on one characteristic property of UJ 
found when simulating the wire breaking: the UJ-matter 
deforms very easily. This ability to deform allows the formation of
the cylinders of magic radii glued to the leads. The radius jumps
from one magic radius to the next through an abrupt charge
reorganization. The CDS's of about two or eight electrons appear before the 
last charge reorganizations and the wire breaking.
If there is enough  UJ between the leads suspended long thin cylinders
appear and in the last steps they alternate with chains of CDS's producing 
a very extended elongation process.
Here we want to underline that the CDS formation is a process
different from the stability of a uniform cylindrical wire against the
formation of a chain of spheres studied in the previous Section
\ref{stability}. In that section, the quantum-mechanical shell structure may conserve the cylindrical structure
which is not classically stable whereas now the quantum-mechanical shell-structure 
effect destroys the classical catenoid-type of solution producing a CDS in the constriction.

In Fig. \ref{force} we show the conductance, the effective radius, and
the elongation force as a function of the elongation for two different
wires. The main figures correspond to an initial configuration with the
radius of $20.7\ a_0$ and 60 electrons in the UJ-constriction. The
insets display the results for a wire with an initial radius of 
10.7 $a_0$ and eight electrons in the constriction. The electron density 
of the latter wire is plotted in Fig. \ref{rho} at certain
elongation stages.

The conductance is calculated with the adiabatic and
semiclassical approximation used by Brandbyge {\it et
al.} \cite{brandbyge}. The constriction is divided into transversal
slices. Then for each slice a uniform wire with the radial extent
of the slice is built and the energy eigenvalues of the subband
bottoms are calculated for this slice. The subband bottoms give effective
potentials along the wire axis. If we look at the dependence of one of them
on the position we see that it raises at the constriction due to the
strong confinement (see Fig. \ref{pot}). The electrons in this subband at the
Fermi energy of the leads have to overcome this barrier in order to carry 
current. To evaluate the transmission probability of the electrons at the Fermi
level through the barrier the semi-classical WKB formula is
used.

The properties of the nanowires have been demonstrated to be dominated
by the narrowest part of the constriction. Therefore we calculate an
effective radius by evaluating the 
electron density per unit length at the middle of the wire.
% its minimum in the constriction (see the density in Fig.
% \ref{rho}(a)).
% If there are two electron density minima in the
% constriction (as in Figs. \ref{rho}(b) and \ref{rho}(d))
% we use the mean electron density per unit length between the two minima. 
It is obtained with the value of the bulk electron density (corresponding to $r_s=4.176\
a_0$).  Fig. \ref{force}(b)
shows the effective radius as a function of the elongation of the wire.
The plateaus or shoulders are in good
coincidence with the infinite wire magic radii of $10.3\
a_0$, $7.3\ a_0$ and $4.3\ a_0$. For the larger wire shown in
Fig. \ref{force}(b) also a small kink can be seen at $\Delta
L=15.5\ a_0$, which corresponds to the magic radius of $13.6\ a_0$. 
Wider magic radii do not appear because of the beat
region of the supercell structure. In the inset 
% the arrowheads represent 
% the effective radius calculated using the density in the middle of 
% the constriction. In this determination
at the end of the plateaus the effective radius increases when elongating 
the wire due to the CDS 
formation. The sudden decrease of the effective radius, accompanied by a
step in the conductance, is due to the sharp charge rearrangements in
the constriction.

\begin{figure*}
\includegraphics[width=1.8\columnwidth] {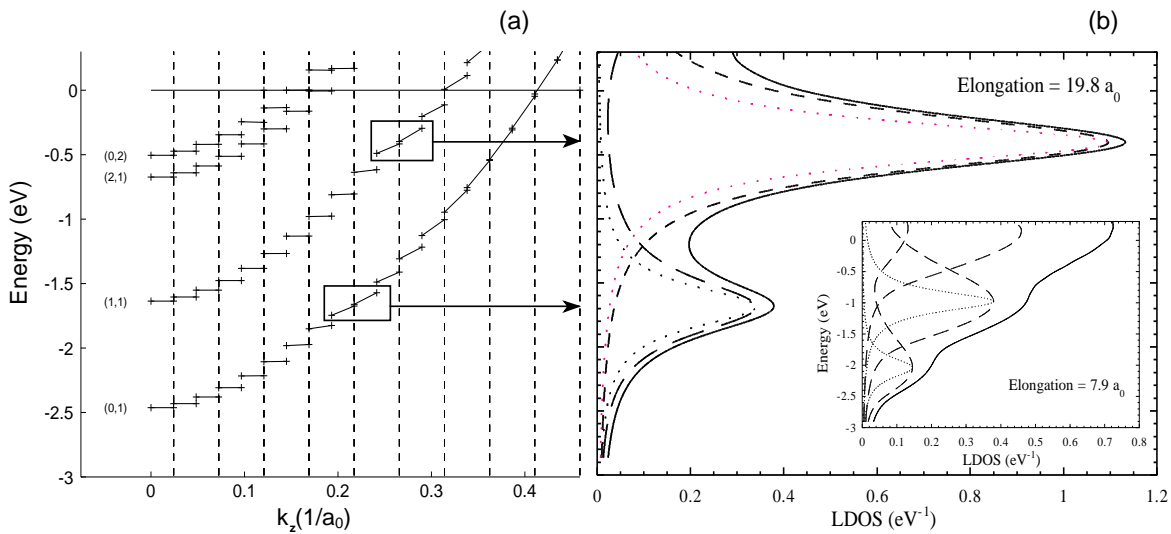}
\caption{\label{band} (a) Electron band
structure of the wire having eight UJ-electrons and  the
elongation of $\Delta L=19.8\ a_0$ (Fig. \ref{rho}(b)).
The vertical dashed lines mark the different
Brillouin zones. The label for each branch represent the
$(m,n)$ subband for the infinite wire. The energy eigenvalues are solved
at two k$_z$-points: at the origin and at the zone boundary of the
supercell Brillouin zone. 
(b) LDOS integrated between the two narrowest points of the 
electron density in Fig. \ref{rho}(b) is displayed. The solid line 
represents the
total ILDOS,  the dashed lines are the contributions to the ILDOS of the
states with m=0 (long-dashed line) and  m=1 (short-dashed line). The
dotted lines show the DOS of hypothetical localized states. The
states marked with squares in the band structure are the main states
contributing to the ILDOS. In  the inset, the analogous plot for the
ILDOS in the constriction of Fig.  \ref{rho}(a) with $\Delta L=7.9\ a_0$ is
shown. The origin of energy is the Fermi level.}
\end{figure*}

The elongation force, shown in Fig \ref{force}(c) is evaluated as the negative derivative of the total energy with respect to the elongation.
%The force is evaluated as the opposite number of the derivative of 
%the total energy with respect 
%to the elongation. 
%Fig. \ref{force}(c) shows the elongation force.
The rearrangement of the wire charge leads to discontinuous upward
steps in the force, while if the radius changes smoothly the force
draws a continuous buckling curve.  At this point we want to point out the
superiority of the UJ model in the force calculation over other jellium models
\cite{yannouleas2,yannouleas,nerea}. In contrast with 
experiments \cite{rubio,rubio-bollinger}, the latter show a continuous
behavior of the force without any steps. Moreover,
for narrow constrictions positive values are obtained when the wire
crosses an unstable zone. Note that in our model the force is always
negative, as observed in the experiments \cite{rubio,rubio-bollinger}
and in atomistic simulations \cite{landman,sorensen,brandbyge,nakamura}. 
Fig. \ref{force} shows clearly that the transport, geometrical and mechanical
properties of the nanowires under elongation are related.
%The relation between transport, geometrical and mechanical
%properties of the nanowires has been revealed as well.

\subsection {Electronic cluster-derived structures}

Let us now analyse more closely the CDS appearing in
Fig. \ref{rho}(b). In order to enlighten the origin of this structure we
plot in Fig. \ref{band}(a) the single-particle energy spectrum of the
wire. The extended zone scheme is used for clarity.
% To perform
% the translation of the eigenenergies out of the first Brillouin zone we
% take into account that the wave functions at the cell boundaries
% should match those at the limit of an infinite wire.
The labels on the left of each branch represent the corresponding 
$(\vert m \vert,n)$ subbands for the infinite
wire. 
%$m$ is the angular momentum quantum number and $n$ is related with the 
%number of nodes of the wave function in the radial direction.
In practice, $n$ is obtained by calculating the number of radial 
nodes at the cell boundaries (See Fig. \ref{sketch}). 
%The m quantum number is the absolute value of states with angular 
%momentum $\pm m$. The $n$ quantum number is related with the number of nodes 
%of the radial part of 
%the infinite cylindrical wire wavefunction. We obtain it by counting the number
%of nodes of the radial part of the wave function at the cell boundaries,
%where it should match those at the limit of an infinite wire.
% Each state has a well-defined $m$ angular quantum
% number and the $n$ quantum number is obtained by counting for the number
% of nodes of the radial part of the wave function at the cell boundaries
% where it should mach those at the limit of an infinite wire. 
The branches have the characteristic parabolic shapes, but they show two different
stages. In the lower part of the parabolic subbands the eigenvalues form
flat plateaus without $k_z$ dispersion. These states correspond
to wavefunctions localized at the leads and they vanish at the center of the 
constriction. 
%with zero probability of presence in the constriction. 
Therefore the (0,2) and (2,1) subbands can not carry current through the 
constriction and they are closed channels. On the other hand, the states 
of the upper part of the (0,1) and (1,1) branches are extended along the 
whole wire and they form a continuous band (with the exception of 
small band gaps). The conductance of the wire is thus $3\ G_0$
due to the extended states of the $(0,1)$ and $(1,1)$ open channels
at Fermi energy. This conclusion is in accordance
with the value obtained with the WKB approach.

% (Martti: How have you calculated ILDOS in practice?? FWHM for the
% peaks corresponding to the energy eigenvalue??)
 
\begin{figure*}
\includegraphics[bb= 40 289 544 729, width=1.8\columnwidth]{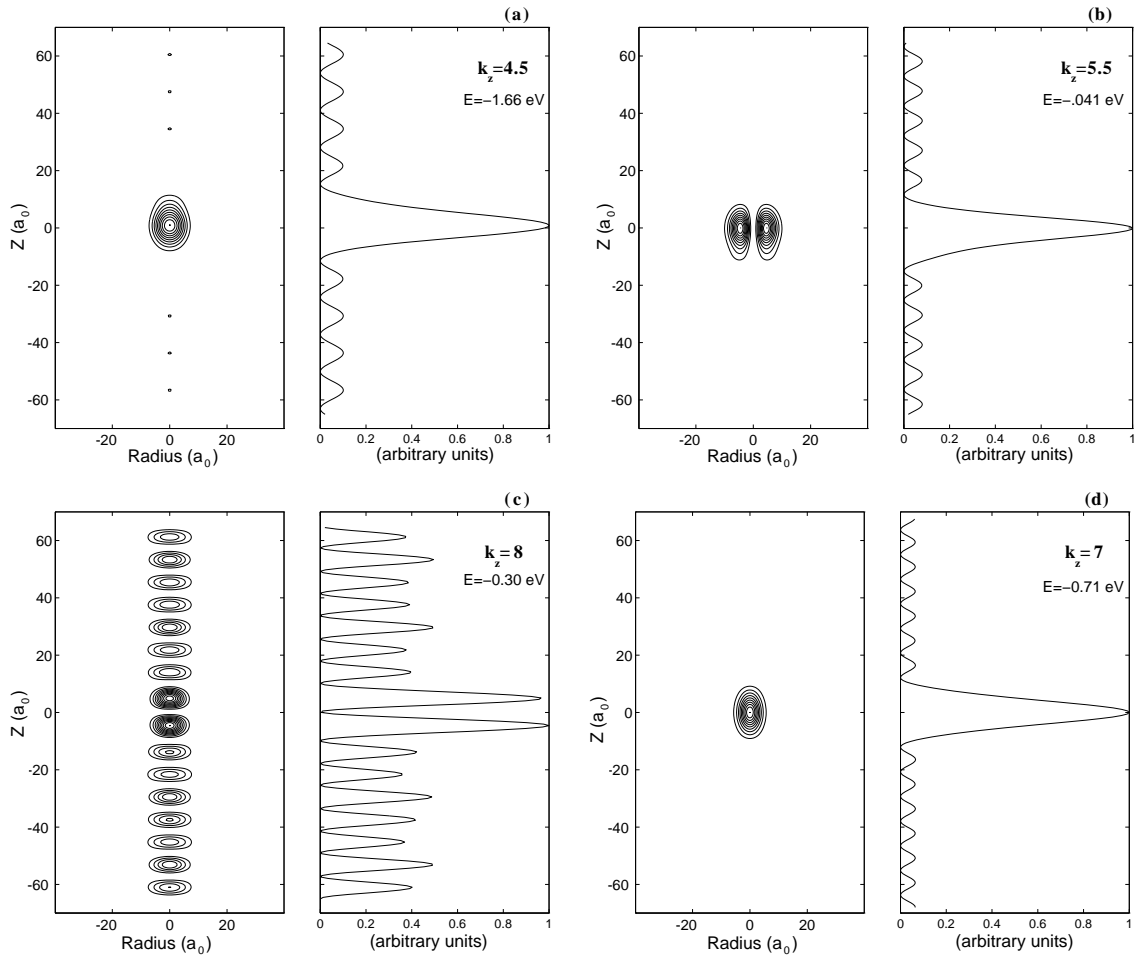}
\caption{\label{waves} (a), (b) and (c): Selected single-electron states 
in the wire having eight UJ-electrons and the elongation of 
$\Delta L=19.8\ a_0$ (Figs. \ref{rho}(b) and \ref{band}). Contour
and profile plots (along the wire through the maximum value) of the
density are shown. Plot (a) corresponds to a $m=0$ resonance state at 
the low energy peak in the ILDOS (Fig. \ref{band}).  Plot (b) corresponds to a
$m=1$ resonance state at high energy. Plot (c) is an extended state
of the $m=0$ subband at the energy of the $m=1$ peak in the ILDOS. Plot
(d) is a localized state with $m=0$ corresponding to the elongation of
$\Delta L=25.8\ a_0$ (Fig. \ref{rho}(c)). The contour
spacing is one tenth of the maximum value. The $k_z$
vector is given in Brillouin zone units $(\frac{2\pi}{L_{\rm cell}})$.}
\end{figure*}
 
In  Fig. \ref{band}(b) we plot the integrated local density of states
(ILDOS) in the constriction for the band structure of
Fig. \ref{band}(a). It is calculated by integrating the local density of states 
(LDOS) over the space between the two narrowest parts in the electron density 
in Fig. \ref{rho}(b). The LDOS itself is obtained by substituting the 
discrete energy levels with Lorenzians of width (FWHM) of 0.4 eV and 
weighting them by the local
probability amplitudes of the states in question. The ILDOS has two clear 
peaks, and when decomposing it we can see that the lower and the higher peak
have the $m=0$ and $m=1$ character, respectively. The
contribution of the $m=2$ states is negligible. 
% For a localized state
% the density of states (DOS) is a Dirac delta function, but it does not
% appear like a delta because in the calculation the DOS is convoluted
% with a smooth function. 
The two ILDOS peaks can be fitted by two energy levels convoluted with 
the same Lorenzian as the eigenlevels in the LDOS calculation. The
resulting resonance peaks are shown in Fig. \ref{band}(b) by dotted
lines. 
%The curves corresponding to hypothetical localized states (delta functions 
%convoluted with the same Lorenzian function which is used 
%when calculating the LDOS) have been plotted with dotted lines. 
The positions and the heights of these peaks have been fitted manually. 
The coincidence between the fit 
%the LDOS of the hypothetical localized states 
and the true ILDOS is
remarkable. In the inset of Fig. \ref{band}(b) we plot the
LDOS integrated between the leads for the electron density showed in
Fig. \ref{rho}(a) having no CDS. We observe that the ILDOS is
much smoother and it is similar to the DOS of an infinite wire with 
delocalized states. The different $m$ contributions cannot be fitted
by single resonance peaks as shown by the dotted peak for the 
$m=0$ and $m=1$ contributions. Moreover, the inset shows that
%modes do
%not fit well the DOS of a hypotetical localized state (drawn with dotted lines 
%in the figure).
%In addition, in the inset 
the $m$-decomposed peaks are slightly  asymmetric with a tail on the 
high energy side. These tails, which are 
not observable in the main figure in which the CDS appears, are
due to the $\sqrt \epsilon$-dependence of the subband peaks in the DOS for
infinite wires.

The ILDOS analysis suggests that in the energy subbands or branches, at
the transition points from states localized in the leads to states extended
across the whole wire (see Fig. \ref{band}(a)), rather localized 
resonance states appear
% (Martti: should we speak here
% and below rather about resonance states than localized states??) 
in the constriction. To clarify this point we plot selected states 
at the ILDOS peak energies in Fig. \ref{waves}. Figs. \ref{waves}(a)
and \ref{waves}(b) show clearly the localized character of the
wavefunctions in the constriction at these energies. 
The state in Fig. \ref{waves}(a) can be identified as the $1s$
orbital of an eight-electron cluster. 
%Speaking in terms
%of orbitals and attending to the $m=0$ quantum number of
%Fig. \ref{waves}(a) wavefunction, it can be identified with the $1s$
%orbital of the eight-electron cluster. 
The second well-localized state, Fig.
\ref{waves}(b), has $m=1$. Therefore it is doubly degenerate and it is
identified as the $p_{xy}$ orbital. At about the energy of this
$p_{xy}$ orbital a $p_z$ orbital (directed along the wire axis) should appear
in the $m=0$ branch  in order to complete the eight-electron cluster.
% , but we can not think in the
% same terms about the localization of the Fig. \ref{waves}(c)
% wavefunction, and it has to be considered extended throughout the wire.
However, we do not find such a state with a strong localization in the
constriction. As shown in Fig. \ref{waves}(c) the $p_z$ type states are 
much more delocalized than the $p_{xy}$ resonance states. The difference
reflects the fact that due to the orientation the interaction of the 
cluster $p_z$ orbital with the lead states is much stronger than that
of the $p_{xy}$ orbital. 
%Although the interaction with the leads could shift the energy
%of this orbital, we have checked the absence of this $p_z$ character
%orbital, been wavefunctions around this energy extended as the plotted 
%in Fig. \ref{waves}(c).
% Then, we think that due to its orientation along the $z$ axis
%it interacts more strongly with the leads than the other orbitals
%becoming an extended state, and only has a very weak localized
%character. 
The absence of a well-localized $p_z$ orbital explains the clearly 
non-spherical shape of the embedded cluster in the electron
density plot of Fig. \ref{rho}(b), and also the finding that
there are only 7.1 electrons in the constriction between the
two narrowest cross sections. 
% In the region where the ILDOS has been integrated there are 7.1 electrons.
% The fact that there are 7.1
% electrons between the electron density narrowings also supports the
% hypothesis that the embedded cluster is related to the eight-electron
% cluster. The difference is that the two $p_z$ states are only weakly localized
% in the constriction. We want to remind that all the extended electrons
% contribute to the 7.1 electrons of the constriction and the
% electron counting is only a qualitative argument.

\begin{figure}
\includegraphics[ width=\columnwidth]{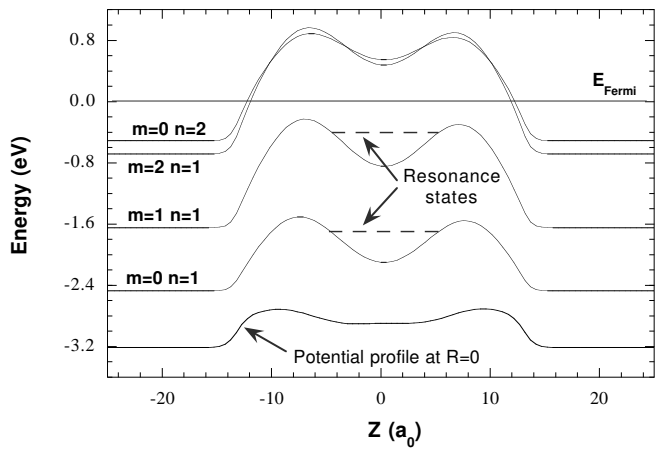}
\caption{\label{pot} 
Effective $z$-dependent potentials for the different $(m,n)$ channels. 
The wire has eight UJ-electrons and the elongation of 
$\Delta L=19.8\ a_0$ (Figs. \ref{rho}(b) and \ref{band}).
The dashed lines in the potential wells are drawn at the energies
of the resonance states. The lowest line is the
bare Kohn-Sham potential at $R=0$.}
\end {figure}

Fig. \ref{waves}(d) shows a well-localized state for the wire with eight 
UJ-electrons at the elongation of $\Delta L=25.8\ a_0$ (Fig \ref{rho}(d)). 
The state can be identified as the $1s$ orbital of
a 2 electron cluster glued to the leads. There are 1.8
electrons between the two narrowest cross sections of the constriction 
supporting the assumption that this state is related to a 
two-electron cluster.

The states in Fig. \ref{waves} have always a wavy, non-decaying background.
This is a characteristic of resonance states; truly localized
states would decay exponentially. The wavy
background corresponds to the wave function of the leads (plane wave)
at the energy, which matches with that of the cluster state. 
To check this assumption we realize that
the wavelength of the plane-wave background corresponds to the $k_z$
quantum number in the extended zone scheme: 
There are indeed two maxima in the modulus of the
wave function per every Brillouin Zone unit of $k_z$
(see the labels of each wave function).

%Note that the  wavefunctions for the localized states have
%always a wavy background, which would be expected to decay
%exponentially if it belonged to a truly localized wavefunction. This wavy
%background corresponds to the wavefunction of the leads (plane wave)
%at this energy, which matches the localized wavefunction at the
%constriction. To hold this assumption we have to realize first that
%the wavelength of this plane-wave background corresponds to the $k_z$
%quantum number in the extended zone scheme (see labels of each
%wavefunction): for each $k_z$ Brillouin Zone unit the modulus of the
%wave function has to show two maxima in the cell. From this point of 
%view these localized states can also be understood as resonance 
%states of the lead wave functions with in the constriction.
% Furthermore, we have
% forced the localized states to disappear partially modifying the
% potential by hand and we observe that the peak in the constriction
% decreases at the same time that the wavy background increases; in the
% limit it will eventually become an extended wire state.

The existence of resonance states is related to the shape of the
self-consistent potential having a small potential well in the
nanoconstriction. To point out
how this potential can admit a resonance state we show in
Fig. \ref{pot} the effective potential for states with
different $(m,n)$ quantum numbers, calculated within the adiabatic
approximation for the wire with eight UJ-electrons and the 
elongation of $\Delta L=19.8\ a_0$ (Fig. \ref{rho}(b)). 
We see that electrons at the constriction
feel the existence of a potential well. We plot the energies 
corresponding to the ILDOS peaks with dashed lines and note that
they lie exactly in the potential wells, where the resonances 
situate. It is also evident that an occupied resonant $p_z$ state 
does not occur because its energy eigenvalue should be well
above the effective potential of the $(0,1)$ branch and
because the potential well of the $(0,2)$ branch is above the
Fermi level. In addition, by the help of Fig. \ref{pot} we can explain 
the different parts of the electron
energy bands in Fig. \ref{band}(a). The states with energies above the
effective potential maxima are extended along the whole wire. These are the 
current-carrying states of each branch. The states below
the potential minimum of the constriction are trapped in the leads,
corresponding to flat plateaus in the lower part of the energy
branches (Fig. \ref{band}(a)). Finally, between the potential maxima
and the local minimum in the center we find resonant states 
which are enhanced at the constriction although they continue as
plane waves in the leads.

\section{\label{Conclusions}Conclusions}

We studied the stability of nanowires and the nanowire breaking process
performing self-consistent calculations within the ultimate jellium model. In
the model, electrons and positive background charge acquire the optimal 
density minimizing the total energy. The model enables thus studies of
shape-dependent properties of nanoscopic systems such as quantum dots
or, as in the present work, quantum wires. The model advocates the idea
that the electronic structure determines via the shell structure the 
geometry and ionic structure also in a partially confined system.

First we analysed the stability of infinite periodic quantum wires
pointing out the ability of the electronic band structure to stabilize
the nanowires at magic radii, {\em i.e.} any small deformation of the
nanowire along the $z$ axis always increases the energy. At the
unstable radii corresponding to maximum values of the energy oscillations 
the wire is uniform up to a critical value of the unit cell length. 
The critical values found are close to the classical value of
$L_{\rm cell}/R=4.5$. Above this limit the local energy minimum disappears
and a deformation of the wire lowers the total energy.

Then we investigated the elongation process of finite nanowires supported 
by leads. The elongation force, conductance and effective radius of the
constriction were calculated simultaneously. The importance of
the charge relaxation, in order to obtain results in agreement with
experiments, was shown, {\em e.g.}, in the case of the elongation
force. The ability of the ultimate jellium (electron density) to acquire 
the optimal shape allows the formation of CDS's showing the importance 
of electron states in the formation of these structures. 
The related resonance states and their
origin was also shown.  We found CDS's that can be linked with
the eight- and two-electron free-standing clusters. 

\begin{acknowledgments}
One of the authors (E.O.) acknowledges the Spanish Ministerio de
Ciencia y Tecnolog\'{\i}a for financial support under the project
PB98-0870-C02 and the Laboratory of Physics of the HUT for the kind
hospitality. T.T. acknowledges financial support from the 
Vilho, Yrj\"o and Kalle V\"ais\"al\"a foundation.
This work has also been supported by the Academy of
Finland through its Centre of Excellence Program (2000-2005).
\end{acknowledgments}


\bibliography{bibliografia}

\end{document}